\documentclass[aps,nofootinbib,showpacs,preprintnumbers,twocolumn,prb,superscriptaddress]{revtex4-1}

\usepackage{amsmath,amssymb}
\usepackage{graphicx}
\usepackage{bm}
\usepackage{color}

\begin{document}

\title{Thermal phase transitions in a honeycomb lattice gas with three-body interactions}

\author{Maximilian Loh\"ofer}
\affiliation{Institut f\"ur Theoretische Festk\"orperphysik, JARA-FIT and JARA-HPC, RWTH Aachen University, 52056 Aachen, Germany}
\author{Lars Bonnes}
\affiliation{Institute for Theoretical Physics, University of Innsbruck, 6020 Innsbruck, Austria}
\author{Stefan Wessel}
\affiliation{Institut f\"ur Theoretische Festk\"orperphysik, JARA-FIT and JARA-HPC, RWTH Aachen University, 52056 Aachen, Germany}

\begin{abstract}
We study the thermal phase transitions in a classical (hard-core) lattice gas model with nearest-neighbor three-body interactions on the honeycomb lattice, based on parallel tempering Monte Carlo simulations.  This system realizes  incompressible low-temperature phases at fractional fillings of $9/16$, $5/8$ and $3/4$ that were identified in a previous study of  a related quantum model. In particular, both the $9/16$ and the $5/8$ phase exhibit an extensive ground state degeneracy reflecting the frustrated nature of the three-body interactions  on the honeycomb lattice. The thermal melting of the $9/16$ phase is found to be a first-order, discontinuous phase transition. On the other hand, from the thermodynamic behavior we obtain indications for a four-states Potts-model thermal transition out of the $5/8$ phase. Employing an exact mapping to a hard-core dimer model on an embedded honeycomb super-lattice, we find that this thermal Potts-model transition relates to the selection of one out of four extensive sectors within the  low-energy manifold of the $5/8$ phase.
\end{abstract}

\pacs{
64.60.De,
 05.50.+q,
 75.40.Mg
}
\maketitle

\section{Introduction}
Classical lattice models with three-body interactions have been intensively studied in the past -- a rather prominent example is the exact solution due to Baxter and Wu of an Ising-like  model with three-body interactions on the triangular lattice~\cite{baxter73}. The Baxter-Wu model exhibits a special four-states Potts model~\cite{wu82} thermal phase transition  without multiplicative logarithmic corrections to the dominant power-law scaling~\cite{nauenberg79,novotny81,schreiber05}. In the low temperature regime, the system selects configurations that relate  to one  of four long-range ordered ground states -- in addition to the fully polarized state,  three other ferrimagnetic states are obtained from the fully polarized state by inverting  all spins on two out of the three sublattices of the triangular lattice. 

Renewed interest into lattice models with three-body interactions emerged more recently due to proposals for realizing such interactions in systems of polar molecules confined to optical lattices with fine-tuned interactions using static electric and microwave fields~\cite{buechler07,schmidt08,capogrosso09,peng09,bonnes10,bonnes11,zhang11}. Within this context, it was observed, that nearest-neighbor three-body interactions on the honeycomb lattice can give rise to complex low-temperature phases already in the classical limit~\cite{bonnes10}. In particular, it was found that the ground state configurations at lattice fillings $n=9/16$ and $5/8$ exhibit an extensive degeneracy, leading to a finite ground state entropy. The details of these phases will be reviewed further below. 

Here, we investigate the nature of the thermal phase transitions out of these low-temperature phases. We find that the thermal melting of the $n=9/16$ phase proceeds via a first-order phase transition. On the other hand, for the $n=5/8$ phase, we obtain from the thermodynamic behavior evidence for  four-states Potts model criticality, without detectable logarithmic corrections to the leading power-law scaling. The Potts-model transition is shown to relate to a separation  of the low-energy manifold into four distinct  sectors, each being equally extensive. This identification can be established through an exact mapping of the ground state sector to states of a classical hard-core dimer model on an embedded honeycomb super-lattice, as detailed below. At the Potts-model transition, the systems spontaneously selects one  of four possible ways of embedding the super-lattice structure onto the underlying lattice. 

Before we present the details of our simulation results on the thermal transitions, we first introduce the model and summarize the results from previous investigations of its ground state properties in the next section. Then, in Sec.~III, we discuss the thermal phase transitions, before we conclude in Sec.~IV.
\begin{figure}[t]
\centering
\includegraphics[width=0.4\columnwidth]{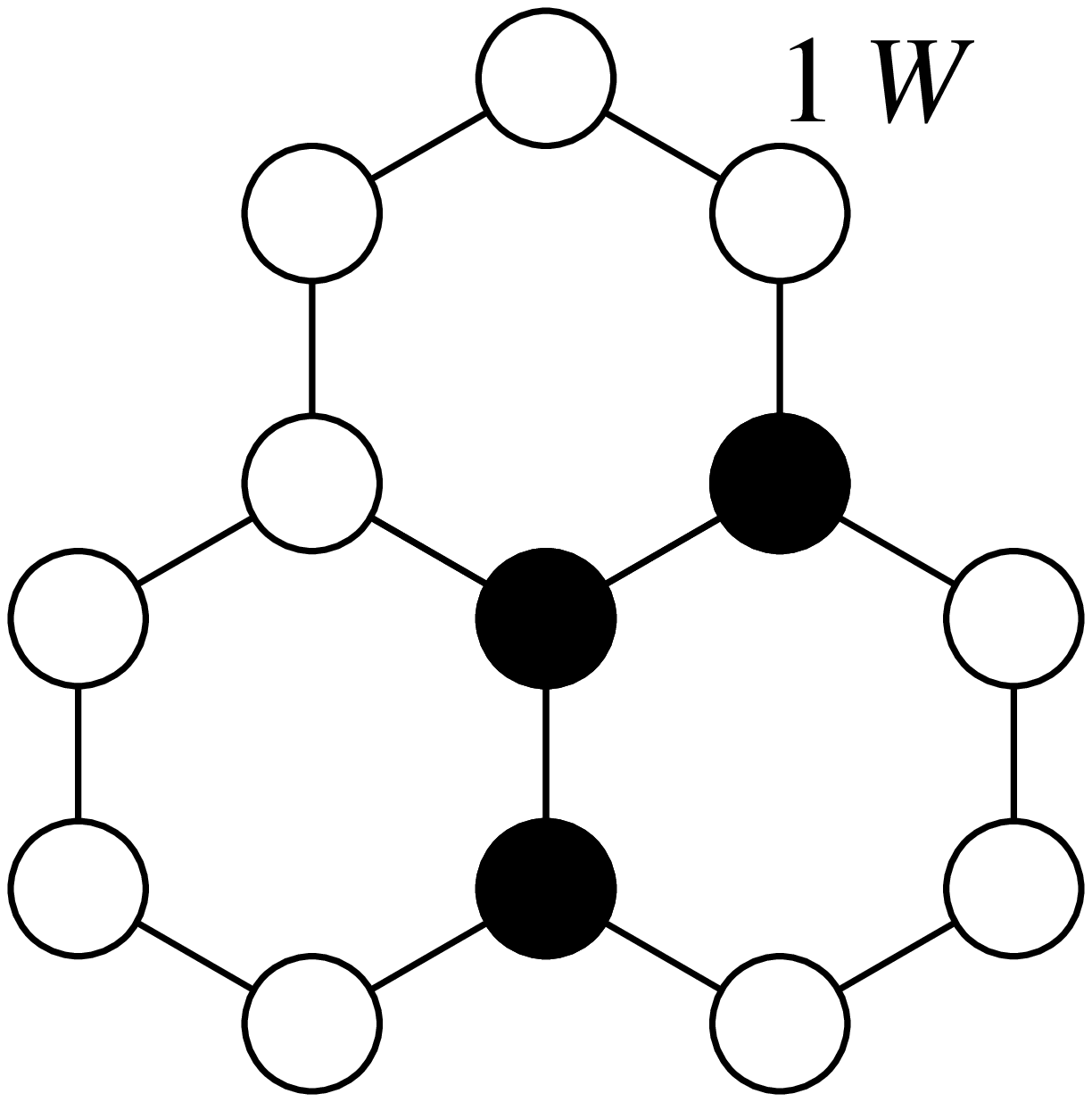}
\includegraphics[width=0.4\columnwidth]{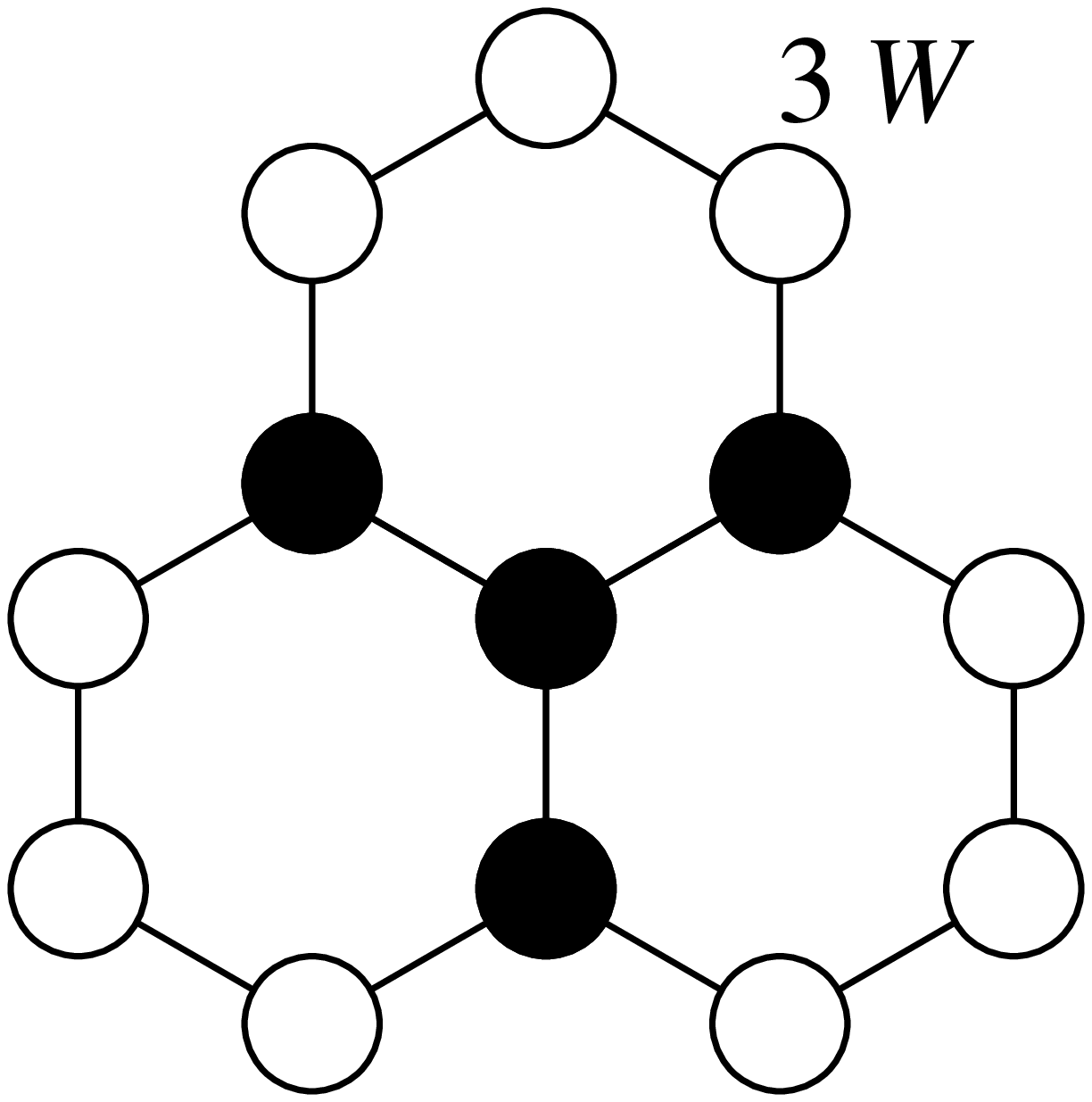}
\caption{Configurations of occupied lattice sites on the honeycomb lattice with an interaction energy 
of $1W$ (left panel) and $3W$ (right panel). 
Open (filled) circles indicate empty (occupied) lattice sites.}
\label{fig:threebody}
\end{figure}

\section{Model and Ground State Phases}
The model that we consider in the following was introduced in Ref.~\onlinecite{bonnes10}, motivated by earlier work on confined gases of polar-molecules on optical lattices~\cite{buechler07}. Namely, we consider here a classical hard-core lattice gas on a honeycomb lattice with a dominant nearest-neighbor three-body interaction, described by
\begin{equation}\label{eq:hamiltonian}
H = W\sum_{\langle i,j,k \rangle}n_i n_j n_k -\mu \sum_i n_i,
\end{equation}
where $n_i\in\{0,1\}$ denotes the local occupation of the lattice site $i$, $\langle i,j,k \rangle$ the sum over all nearest-neighbor triples of sites, and $W>0$ a nearest-neighbor three-body repulsion. Furthermore, the chemical potential $\mu$ allows to control the filling $n$ of the honeycomb lattice. The model can also be expressed in terms of Ising-like variables  $\sigma_i=2n_i-1\in\{-1,1\}$ by the usual mapping from the local occupations.  
The interaction term is illustrated in Fig.~\ref{fig:threebody} for two exemplary configurations. 
The ground state phase diagram of the model as a function of $\mu/W$ can be summarized as follows\cite{bonnes10}: For $\mu<0$ ($\mu>9W$), the lattice is empty (full). In between, the density is pinned to three specific values, each corresponding to a complex structure of the ground state manifold. We now discuss these three different regimes. 

\begin{figure}[t]
 \centering
\includegraphics[width=0.9\columnwidth]{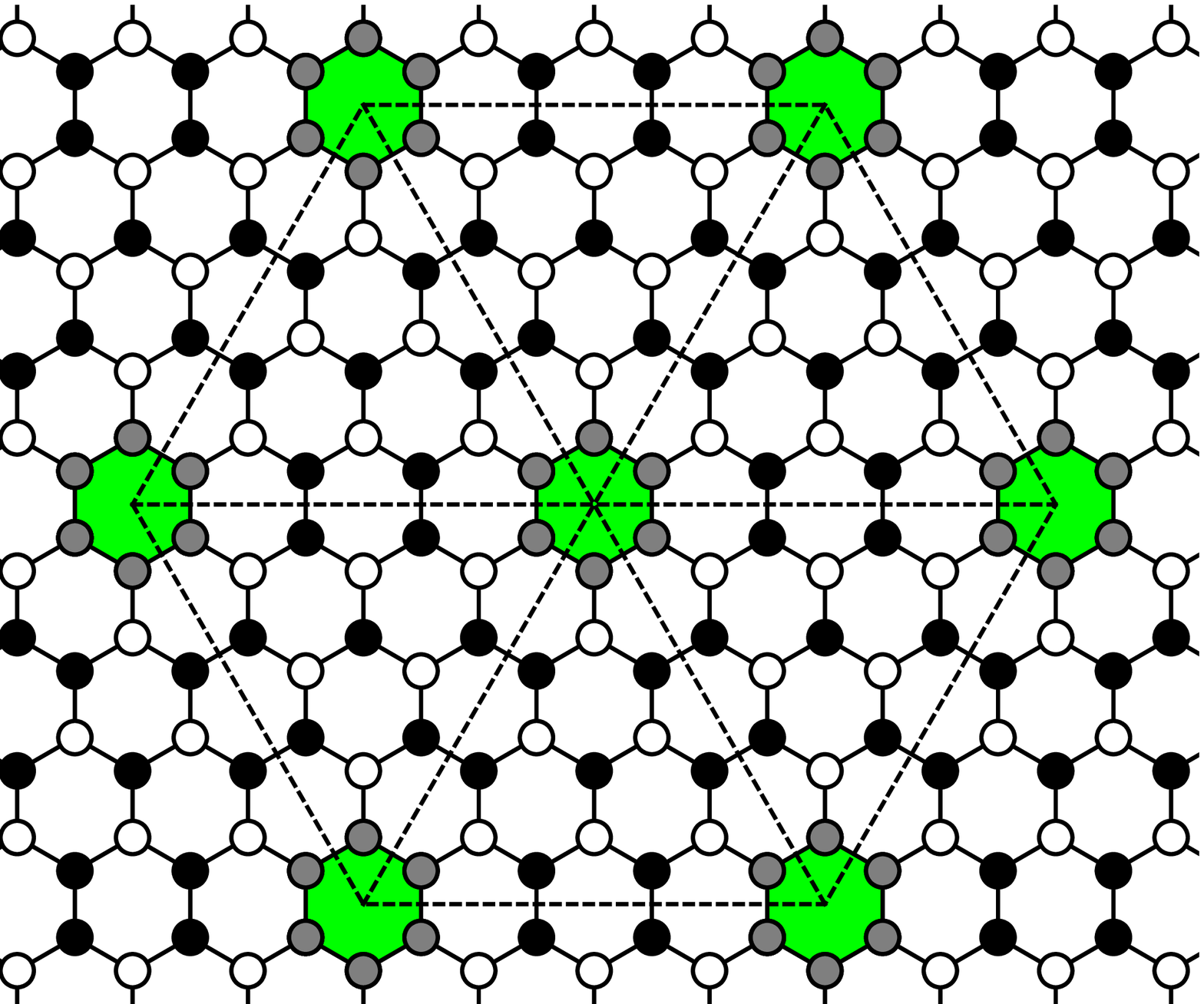}
\includegraphics[width=0.6\columnwidth]{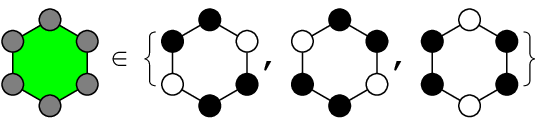}
\caption{(Color online)  Ground state configurations of the $n=9/16$ phase consist of a regular lattice of triangles (upper panel). The configurations around the hexagons at which six such triangles meet can each be taken from the three possibilities shown in the lower panel. Open (filled) circles indicate empty (occupied) lattice sites.}
\label{fig:gs916}
\end{figure}

\subsection{Filling $9/16$ Phase}
\begin{figure}[t]
\centering
$\vcenter{\hbox{\includegraphics[width=0.7\columnwidth]{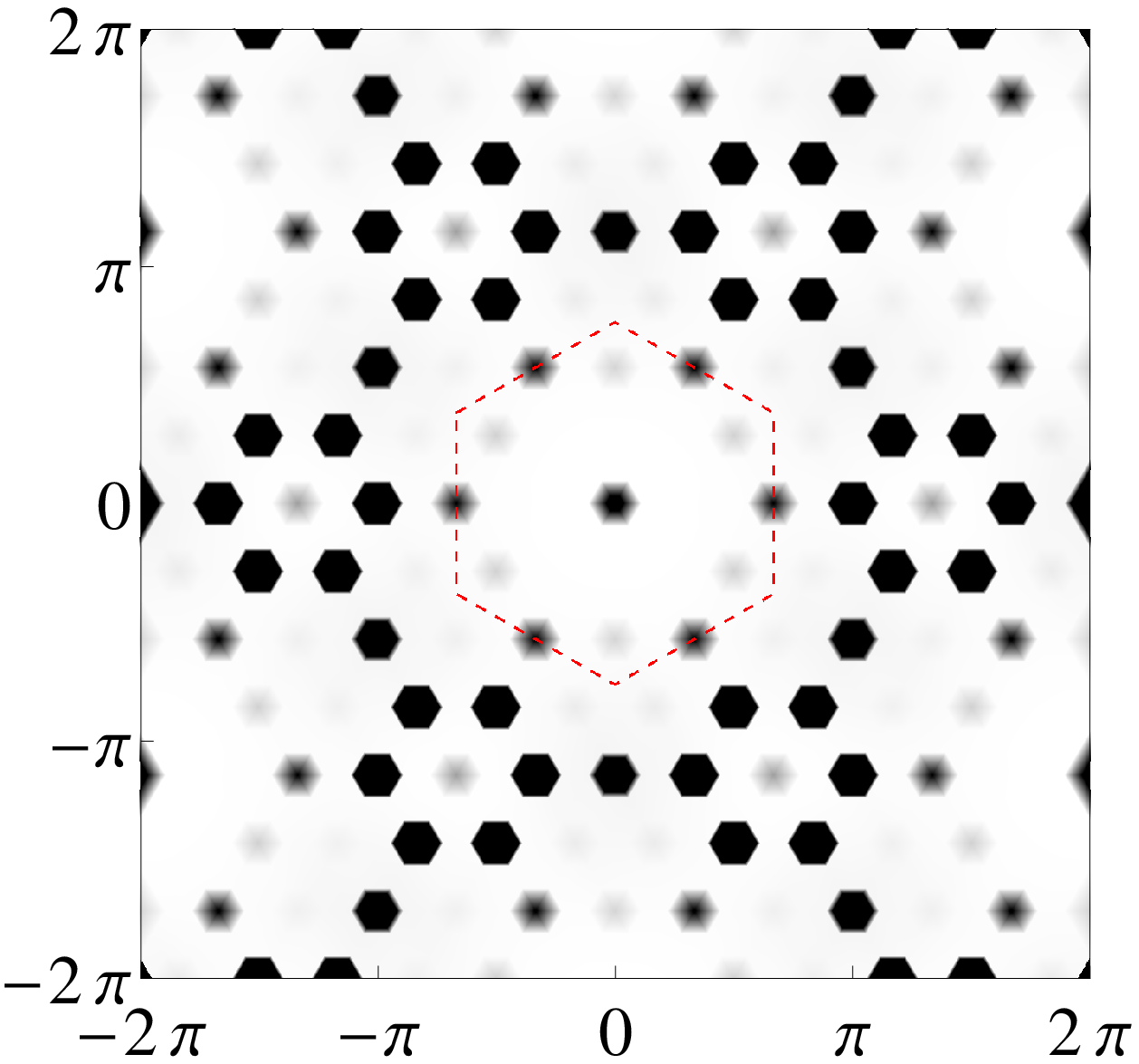}}}$
$\vcenter{\hbox{\includegraphics[width=0.08\columnwidth]{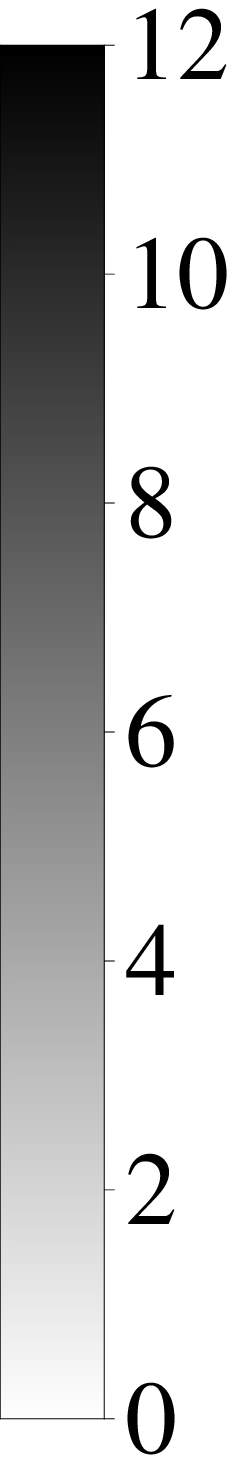}}}$
\caption{(Color online) Structure factor in the ground state for the $n=9/16$ phase. The dashed red hexagon indicates the first Brillouin zone.}
\label{fig:psf916}
\end{figure}

For $0\leq \mu \leq 2W$, the density is fixed to a value of $n=9/16$, and the system realizes an ordered superstructure with a 32-sites unit cell. This  structure is composed out of equilateral triangles, each containing 9 particles and 16 sites. Within each such triangle, the particles are arranged in a staggered pattern. These triangles build up the superstructure in which two neighboring triangles form the new unit cell. 
On hexagons of the underlying honeycomb lattice, where six of these triangles meet, the configuration is not uniquely determined. Instead, each such hexagon can  reside in one out of three different local configurations. The resulting structure of these ground states is illustrated in Fig.~\ref{fig:gs916}. The ground state manifold is thus highly degenerate. In more detail, since each free hexagon is allowed to be in three possible configurations, the ground state degeneracy of a system with $N$ lattice sites scales with $W=3^{N/32}$, leading to the ground state entropy per site of 
\begin{equation}\label{eq:S916}
S=\ln{W}/N=\ln(3)/32\approx 0.034.
\end{equation}
in the thermodynamic limit.
There are no three-body interaction terms penalizing these configurations, so that the ground state energy per site equals $E_0=-9/16\mu$ in this regime. 
The regular arrangement of the  triangular structures leads to a corresponding characteristic peak at $\mathbf{q}_{9/16}=(\pi,0)$ in the  structure factor
\begin{equation}\label{eq:sf}
S(\mathbf{q})=\langle \mathcal{S}(\mathbf{q})\rangle ,\quad 
\mathcal{S}(\mathbf{q}) =\frac{1}{N}\sum_{i,j} e^{i \mathbf{q}\cdot(\mathbf{x}_i-\mathbf{x}_j)} \sigma_i\: \sigma_j ,
\end{equation}
where $\mathbf{x}_i$ denotes the position of the $i$-th lattice site on the honeycomb lattice, in terms of the Ising-like variables $\sigma_i=2n_i-1\in\{-1,1\}$ . The  structure factor in the ground state manifold (obtained from simple sampling from the set of ground state configurations) is shown in Fig.~\ref{fig:psf916}, exhibiting the dominant characteristic peak at $\mathbf{q}=\mathbf{q}_{9/16}$ and the corresponding symmetry-related momenta.

\begin{figure}[t]
\centering    
\includegraphics[width=0.9\columnwidth]{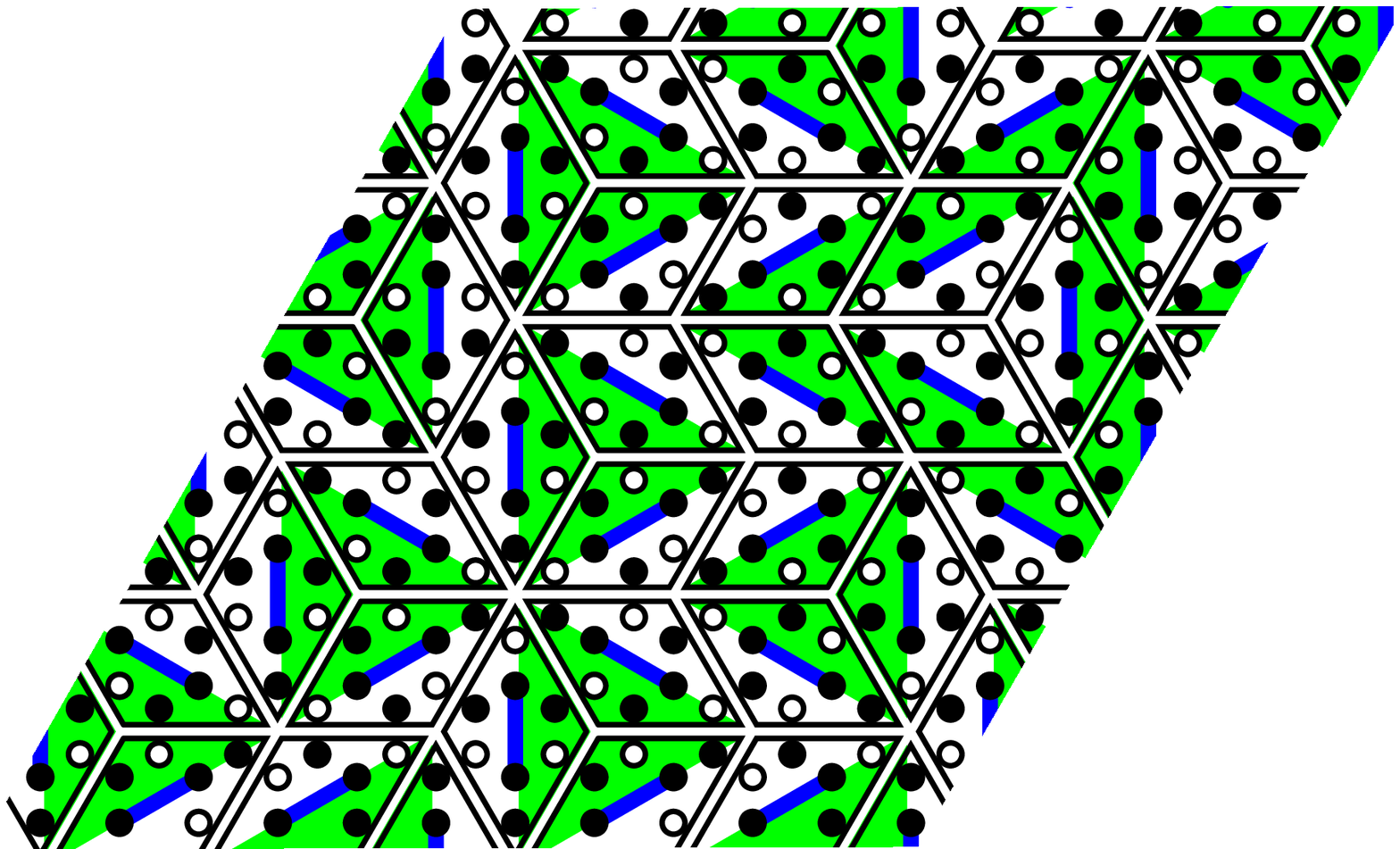}
\includegraphics[width=0.9\columnwidth]{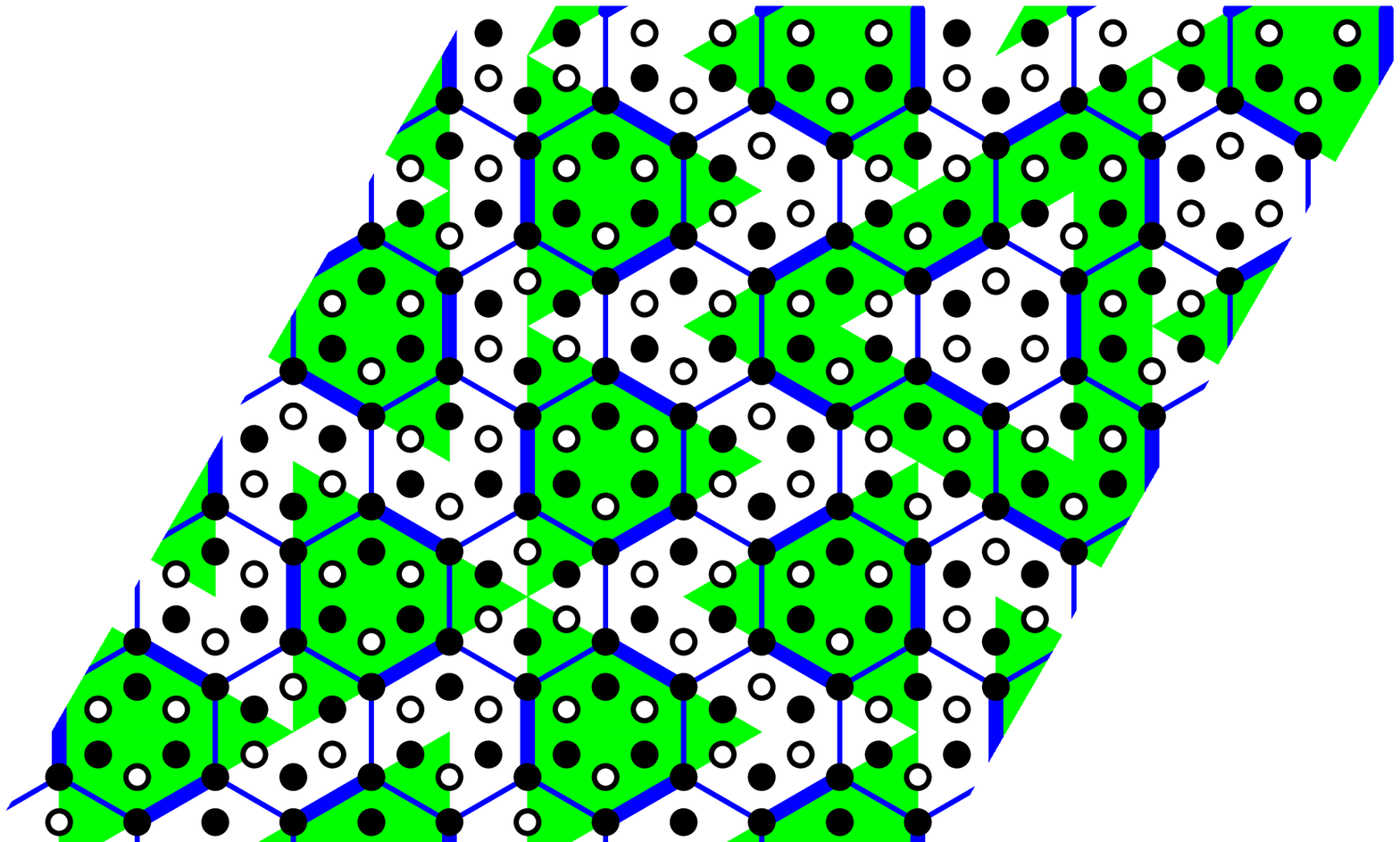}
\caption{(Color online)  Ground state configurations for the $n=5/8$ phase consist of fully packed coverings of the plane by bi-chromatic lozenges, each containing 8 sites with 5 particles, with the constraint, that neighboring lozenges are allowed to touch only along equally colored sides (upper panel). Each such configuration can be mapped onto a hard-core dimer covering of a honeycomb super-lattice, with the central sticks inside each lozenge representing the dimers (lower panel). 
Open (filled) circles indicate empty (occupied) lattice sites. In the lower panel, the honeycomb super-lattice is indicated atop the underlying honeycomb structure (blue lines).}
\label{fig:gs58}
\end{figure}

\subsection{Filling $5/8$ Phase}
\begin{figure}[t]
\centering
$\vcenter{\hbox{\includegraphics[width=0.7\columnwidth]{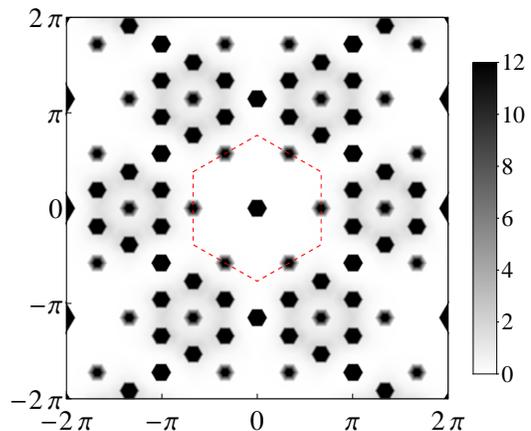}}}$
$\vcenter{\hbox{\includegraphics[width=0.08\columnwidth]{pictures_scale.eps}}}$
\caption{(Color online) Structure factor in the ground state for the $n=5/8$ phase. The dashed red hexagon indicates the first Brillouin zone.}
\label{fig:psf58}
\end{figure}

In the regime $2W\leq \mu \leq 5W$, another incompressible phase is realized, with  filling $n=5/8$. It was observed in Ref.~\onlinecite{bonnes10}, that the configurations which contribute to the ground state manifold in this regime can be related to perfect tilings of the honeycomb lattice by bi-chromatic lozenges containing 5 particles on 8 sites as the elementary units. To ensure the corresponding ground state energy of $E_0=W/8-5\mu/8$ in this regime, the edges of the lozenges are allowed to touch along equally colored sides only. An example of a valid  configuration for the filling $5/8$ phase is shown in the upper panel of Fig.~\ref{fig:gs58}.
If the central two occupied sites in each lozenge are connected by a bold line (cf. Fig.~\ref{fig:gs58}), then each valid lozenge tiling can be identified with a configuration of closed-packed hard-core dimers on a honeycomb super-lattice, as shown in the lower panel of  Fig.~\ref{fig:gs58}. This identification leads directly to the finite ground state entropy $S=0.108$ of the $n=5/8$ phase from that of closed-packed hard-core dimer coverings on the honeycomb super-lattice~\cite{bonnes10}. Note, however, that there is an additional degree of freedom in embedding  the honeycomb  super-lattice atop the underlying honeycomb lattice; in particular, there are four possible ways of realizing this embedding. While this additional factor does not increase the ground state entropy, it nevertheless will be important for understanding the thermal phase transition into the $n=5/8$ phase, as discussed below. Furthermore, while the freedom in tiling the lattice by the lozenge units might appear to eliminate the possibility of long-ranged density-density correlations, this is indeed not the case. In fact, for each given embedding of the honeycomb super-lattice, the central sites of each lozenge are filled for all possible tilings of the lozenges. This induces long-ranged order in the density-density correlation function, and leads to characteristic peaks in the ground state structure factor, shown in  Fig.~\ref{fig:psf58}. We can in particular identify the emergence of these low-energy states by a peak in $S(\mathbf{q})$ at the characteristic wave vector $\mathbf{q}_{5/8}=(2\pi/3,0)$ and its symmetry related equivalents.

\begin{figure}[t]
\centering
\includegraphics[width=0.45\columnwidth]{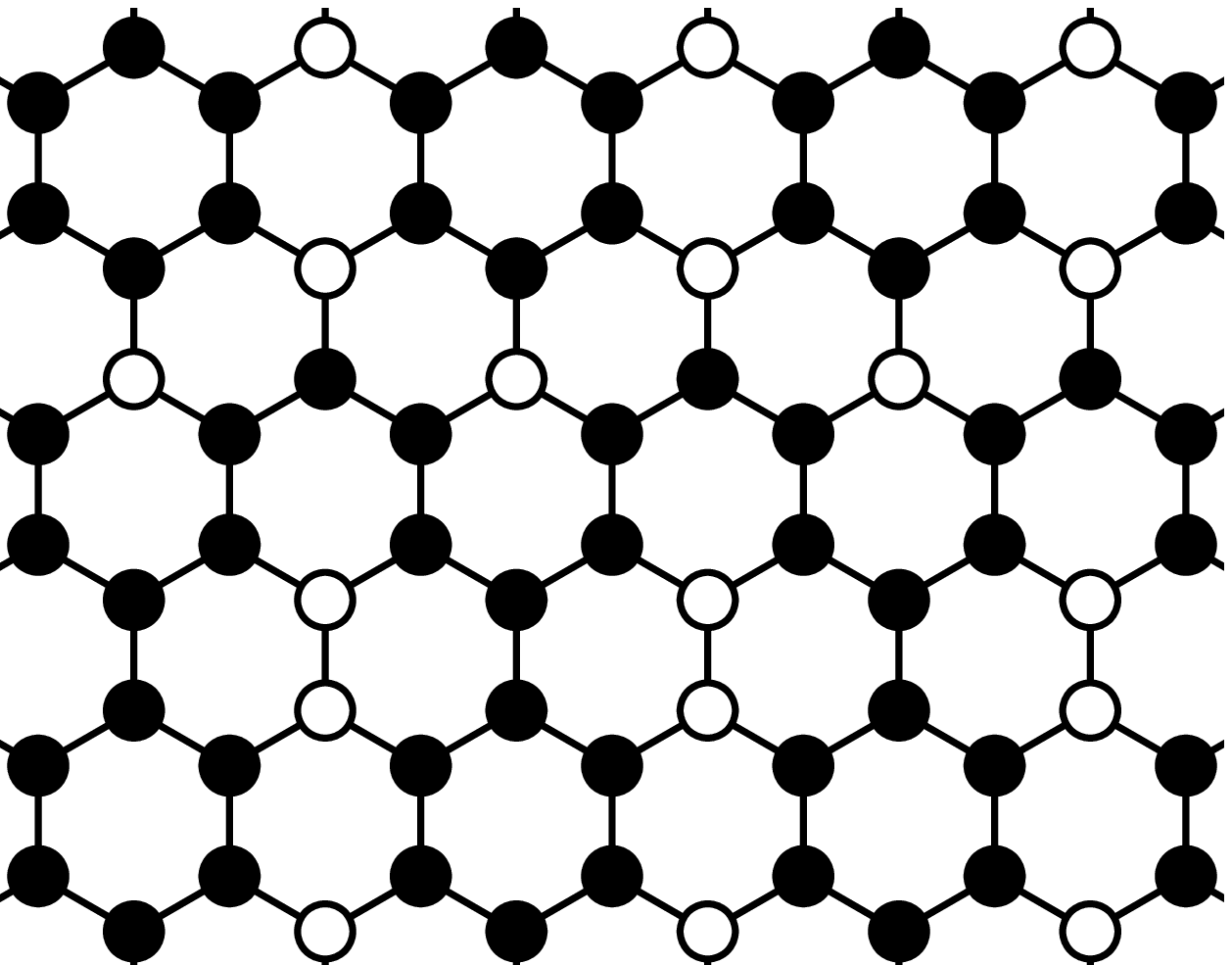}
\includegraphics[width=0.45\columnwidth]{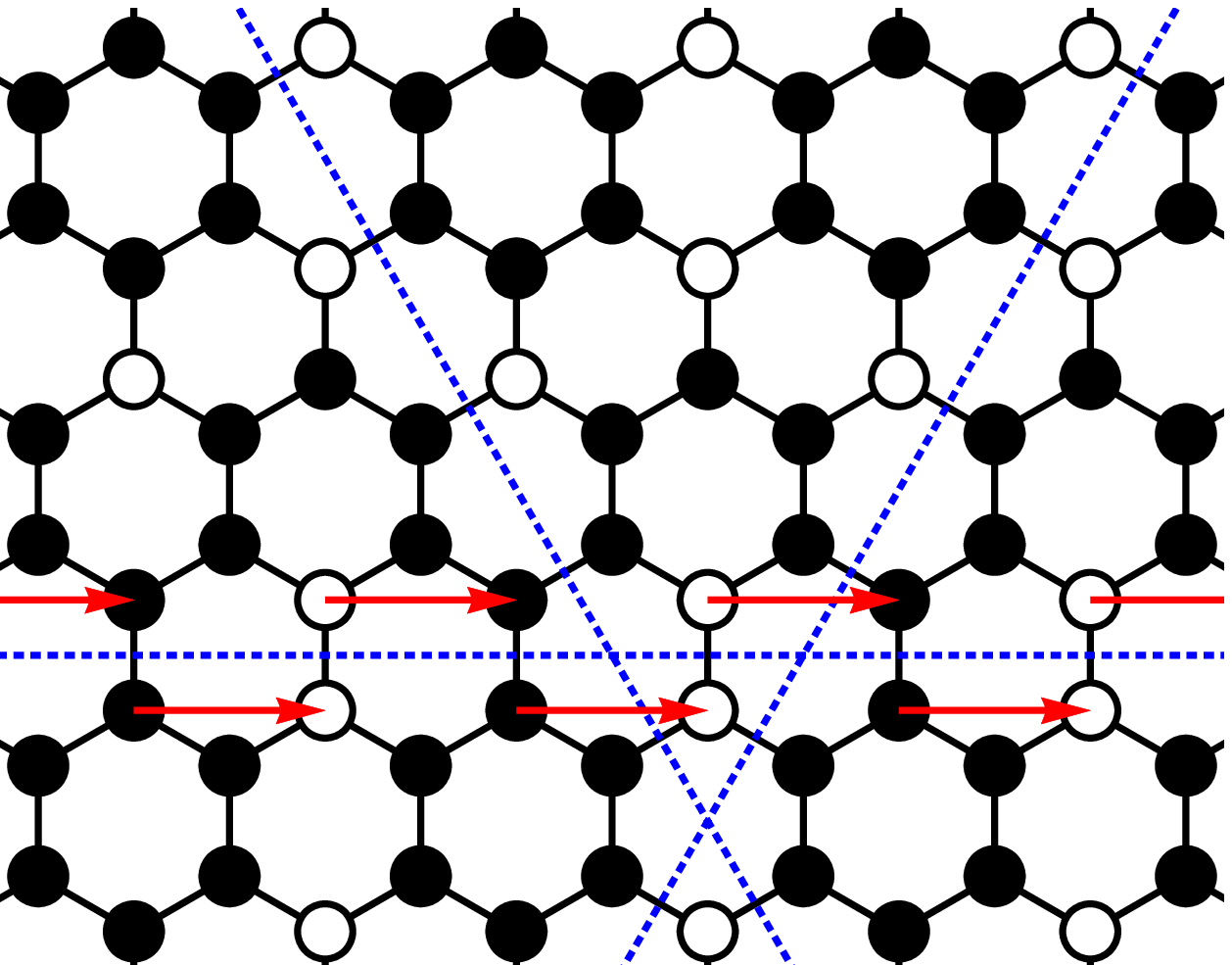}
\caption{(Color online) Ground state configurations in the $n=3/4$ phase consist of a regular super-lattice of  filled hexagons (left panel)
as well as all configurations that result from shifting (arrows) the particles on two parallel lines along one of the lattice directions (dashed lines) (right panel). Open (filled) circles indicate empty (occupied) lattice sites.}
\label{fig:gs34}
\end{figure}

\subsection{Filling $3/4$ Phase}
\begin{figure}[t]
\centering
$\vcenter{\hbox{\includegraphics[width=0.7\columnwidth]{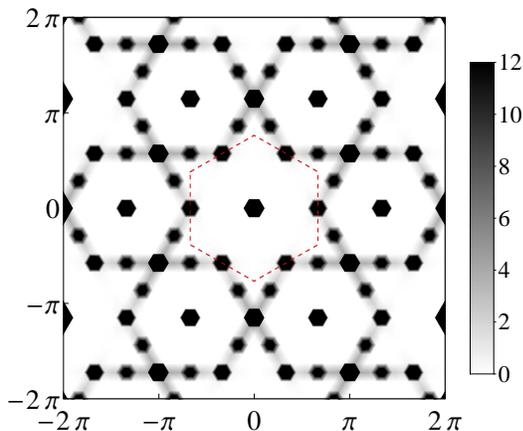}}}$
$\vcenter{\hbox{\includegraphics[width=0.08\columnwidth]{pictures_scale.eps}}}$
\caption{(Color online) Structure factor in the ground state for the $n=3/4$ phase. The  dashed red hexagon indicates the first Brillouin zone.}
\label{fig:psf34}
\end{figure}

Finally, in the range $5W<\mu<9W$, the ground state of the system is composed out of a non-extensive set of configurations with energy $E_0=3/4W-3/4\mu$ and filling $n=3/4$. A specific such state consists of a super-lattice of fully filled hexagons, shown in the upper panel of Fig.~\ref{fig:gs34}. The other states of the ground states manifold can be obtained by applying global shifts of particles along parallel lines throughout the system, as shown in the lower panel of Fig.~\ref{fig:gs34}. Since the number of lines along which such moves can be performed is proportional to the linear system size, the entropy per site scales to zero in the thermodynamic limit~\cite{bonnes10}.  
The restriction of the global moves to align along parallel lines of the honeycomb lattice leads again to a characteristic signature for the $n=3/4$ phase in the structure factor $S(\mathbf{q})$, shown in Fig.~\ref{fig:psf34}, which can be employed to detect the emergence of this specific configurational alignment upon cooling the system, as discussed below.  

\section{Thermal Phase Transitions} 
After having reviewed the system's ground state properties, 
we next investigate the thermal phase transitions out of these phases. We focus here on the filling $9/16$ and the $5/8$ phases, which exhibit an extensive ground state entropy. For both phases, we performed Monte Carlo simulations using local updates, combined with
parallel tempering~\cite{hukushima96} and histogram reweighing~\cite{ferrenberg88,ferrenberg89}. 
For the filling $3/4$ phase, we were not able to perform a similarly systematic study of the thermal phase transition, instead we observed rather strong finite-size shifts in the effective transition temperature (e.g. in the peak position of the specific heat), which indicate that on the finite lattices available to our simulations we were not able to reach sufficiently into the actual transition regime, which would be required in order to perform a controlled finite-size analysis. 

\subsection{Filling $9/16$ Phase}
\begin{figure}[t]
\centering
\includegraphics[width=0.45\columnwidth]{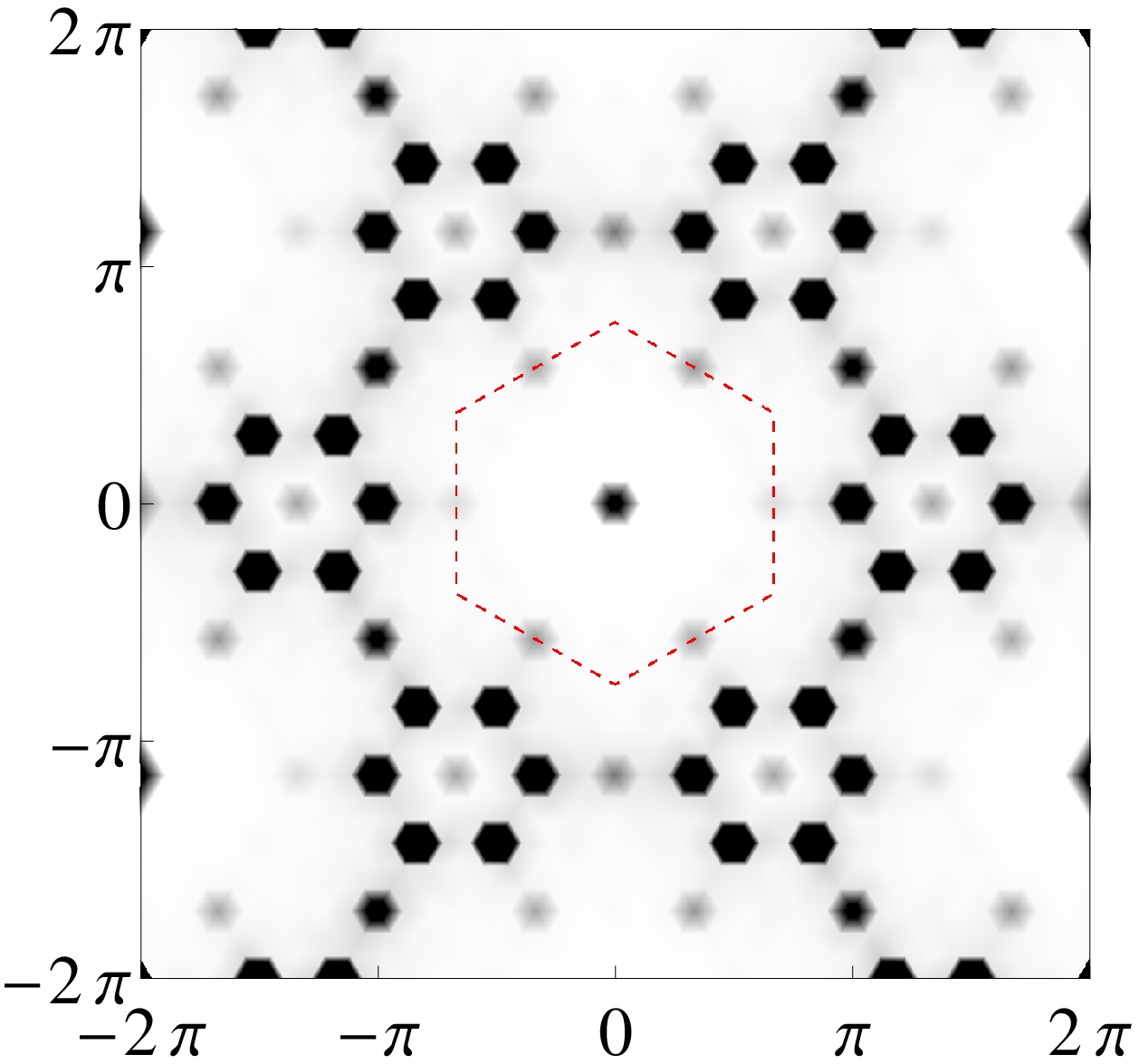}
\includegraphics[width=0.45\columnwidth]{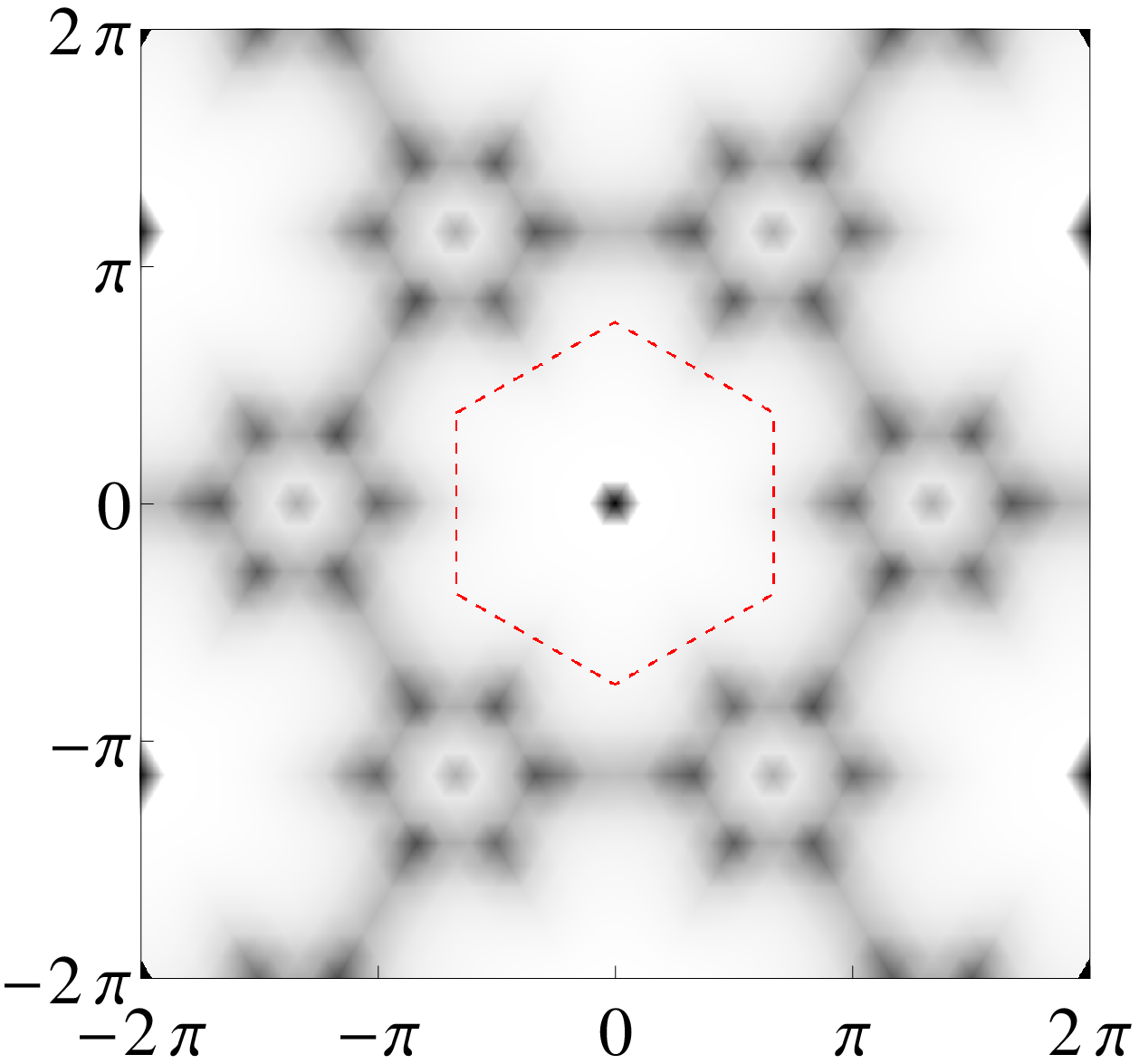}
\caption{(Color online) Structure factor $S(\mathbf{q})$, calculated on a $L = 24$ lattice for temperatures below ($T = 0.07W$, left) and  above the phase transition ($T = 0.24W$, right) to the $n=9/16$ phase at $\mu=1.1W$. The  dashed red hexagons indicate the first Brillouin zone.}
\label{fig:mcsf916}
\end{figure}

\begin{figure}[t]
\centering
\includegraphics[width=\columnwidth]{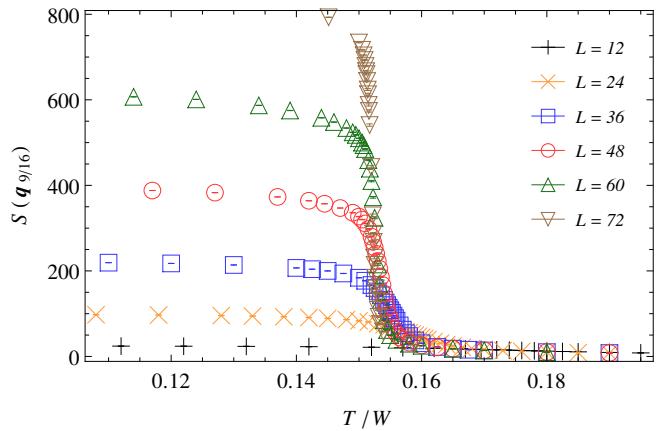}
\caption{(Color online) Structure factor $S(\mathbf{q}_{9/16})$ as a function of temperature $T$ at $\mu=1.1W$ in the transition region to the $n=9/16$ phase for various lattice sizes.}
\label{fig:sf916}
\end{figure}

\begin{figure}[t]
\centering
\includegraphics[width=\columnwidth]{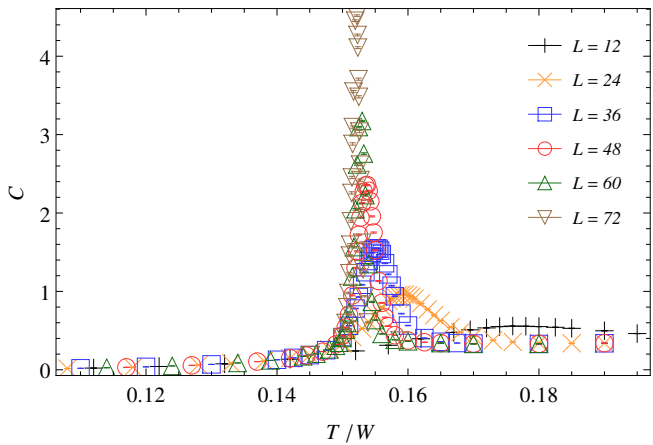}
\caption{(Color online) Specific heat $C$ as a function of temperature $T$ at $\mu=1.1W$ in the  transition region to the $n=9/16$ phase for various lattice sizes.}
\label{fig:C916}
\end{figure}

To study the thermal melting of the structural order in the $n=9/16$-phase, we considered finite systems with periodic boundary conditions and $N=2L^2$ lattice sites, with a linear system size $L$ up to $L=72$, with $L$ an integer multiple of 12, chosen commensurate with the ordering structure. In the following, we present in detail our data for a chemical potential of $\mu=1.1W$, chosen such as to locate the system well inside the $n=9/16$ regime, and  increase the critical temperature scale. 
In Fig.~\ref{fig:mcsf916}, the structure factor $S(\mathbf{q})$ is shown for different values of the temperature $T$, exhibiting at low $T$ the formation of the characteristic structure that is present also in Fig.~\ref{fig:psf916} for the ground state. In more detail, the temperature dependence of the structure factor at the ordering wave vector, $S(\mathbf{q}_{6/16})$, is shown in Fig.~\ref{fig:sf916}.
From this data, the onset of the low-$T$ ordered phase near $T= 0.155W$ can be extracted. 
The phase transition is also monitored by the specific heat $C$, shown in Fig.~\ref{fig:C916}, which exhibits a pronounced peak near $T=0.155W$. 

\begin{figure}[t]
\centering 
\includegraphics[width=\columnwidth]{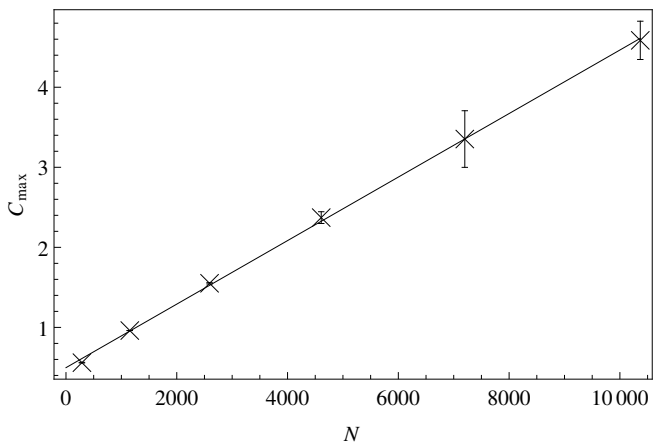}
\caption{Specific heat maximum $C_{max}$ as a function of system size $N=2L^2$ at $\mu=1.1W$ in the  transition region to the $n=9/16$ phase. The line denotes a linear increase with $N$.}
\label{fig:Cmax916}
\end{figure}

\begin{figure}[t]
\centering
\includegraphics[width=\columnwidth]{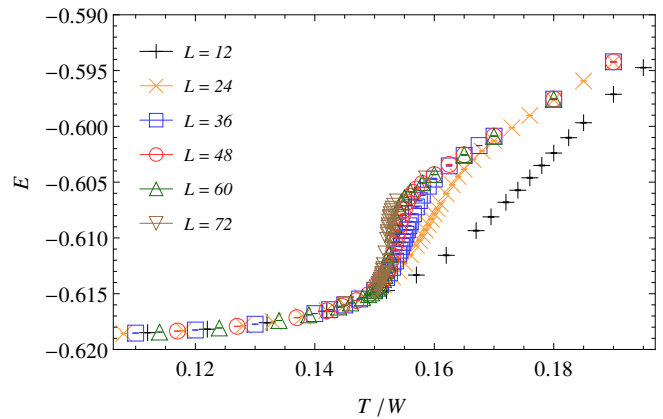}
\caption{(Color online) Internal energy per site $E$ as a function of temperature $T$ at $\mu=1.1W$ in the  transition region to the $n=9/16$ phase for various lattice sizes.}
\label{fig:E916}
\end{figure}

\begin{figure}[t]
\centering
\includegraphics[width=\columnwidth]{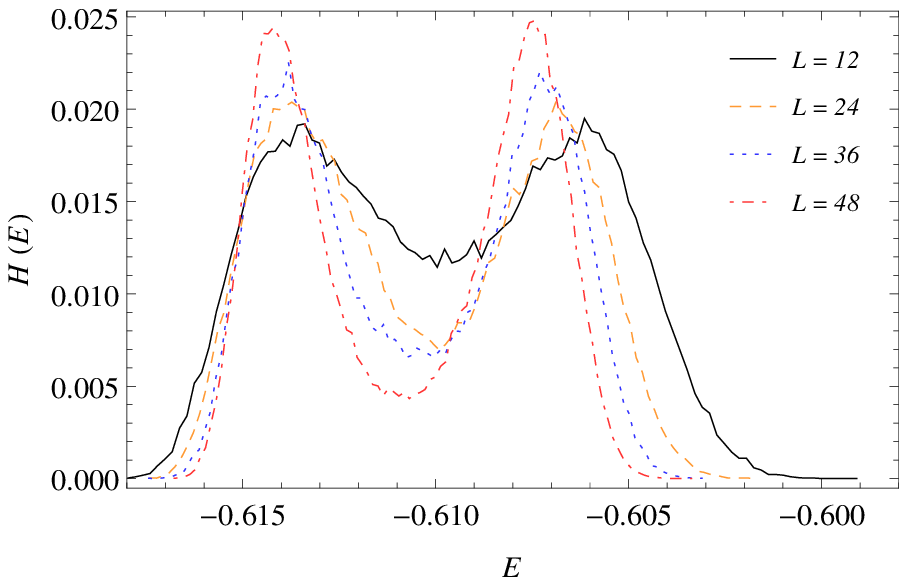}
\caption{(Color online) Histogram of the internal energy per site $H(E)$  near the transition point to the $n=9/16$ phase for various lattice sizes at $\mu=1.1W$. For each system size, $H(E)$ is shown at the temperature for which the high- and the low-energy peak exhibit similar weights.}
\label{fig:HE916}
\end{figure}

Using  histogram reweighing, we extracted the maximum value $C_{max}$ of the specific heat at the peak position, and show the finite-size scaling  of $C_{max}$ in Fig.~\ref{fig:Cmax916}. The observed linear increase of $C_{max}$ with the system size $N$ is indicative of a first-order phase transition. Further support for a discontinuous transition emerges from analyzing the internal energy, shown in  Fig.~\ref{fig:E916}, and in particular from the energy histogram $H(E)$, shown in Fig.~\ref{fig:HE916}. The histograms $H(E)$ have been obtained using histogram reweighing, adapting for each system size the temperature such that the relative weight of the low- and the high-energy peak are of comparable magnitude. The finite-size data of $H(E)$ exhibit an increasing depletion of $H(E)$ between the peak energy positions upon increasing the system size $N$, thus providing another characteristic feature of a first-order phase transition.

\begin{figure}[t]
\centering
\includegraphics[width=\columnwidth]{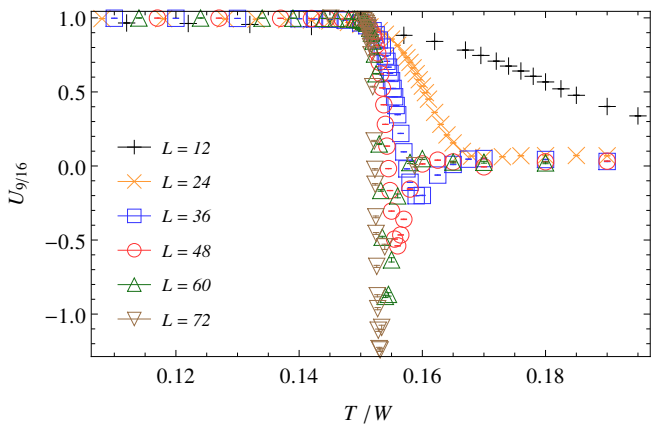}
\caption{(Color online) Binder ratio $U_{9/16}$ as a function of temperature $T$ at $\mu=1.1W$ in the thermal transition region to the $n=9/16$ phase for various lattice sizes.}
\label{fig:U916}
\end{figure}

We also calculated an appropriate Binder cumulant ratio, defined in terms  of the structure factor $S(\mathbf{q}_{6/16})$ as
\begin{equation}\label{eq:binder}
 U_{9/16}=A\left(1-B\frac{\langle \mathcal{S}(\mathbf{q}_{6/16})^2\rangle }{\langle \mathcal{S}(\mathbf{q}_{6/16})\rangle^2}\right),
\end{equation}
to study this phase transition. Here, $A$ and $B$ were chosen such that $U_{9/16}\rightarrow 1$ (0) in the low (high) temperature limit.  
The simulation data for $U_{9/16}$ in the transition region is shown in Fig.~\ref{fig:U916}.
We observe in the finite-size data of $U_{9/16}$ a prominent dip of increasing magnitude upon increasing the system size, which typically accompanies a first-order transition. From the above finite-size analysis, we thus conclude, that the $n=9/16$ phase melts via a direct first-order phase transition.

\subsection{Filling $5/8$ Phase}
\begin{figure}[t]
\centering
\includegraphics[width=0.45\columnwidth]{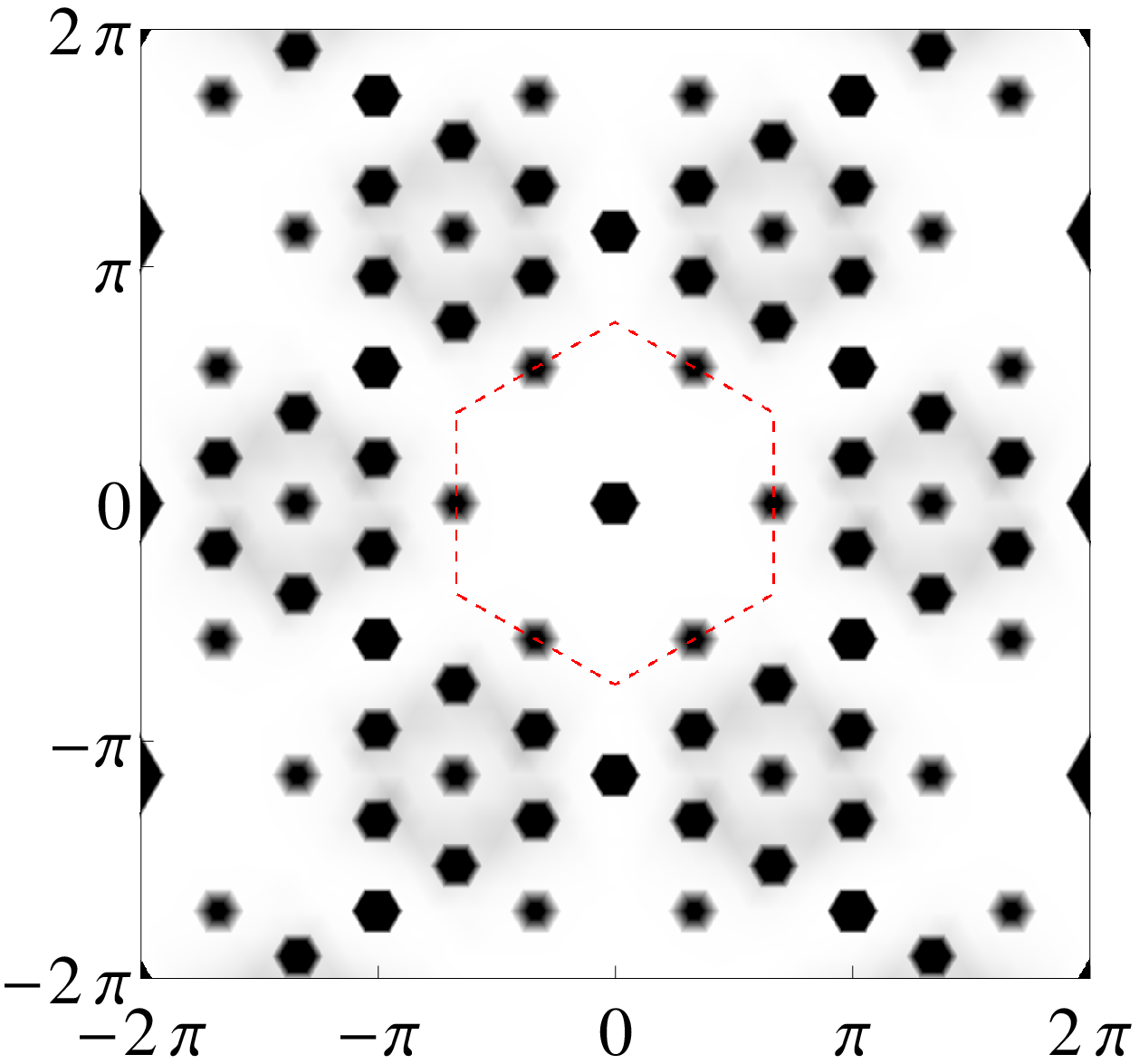}
\includegraphics[width=0.45\columnwidth]{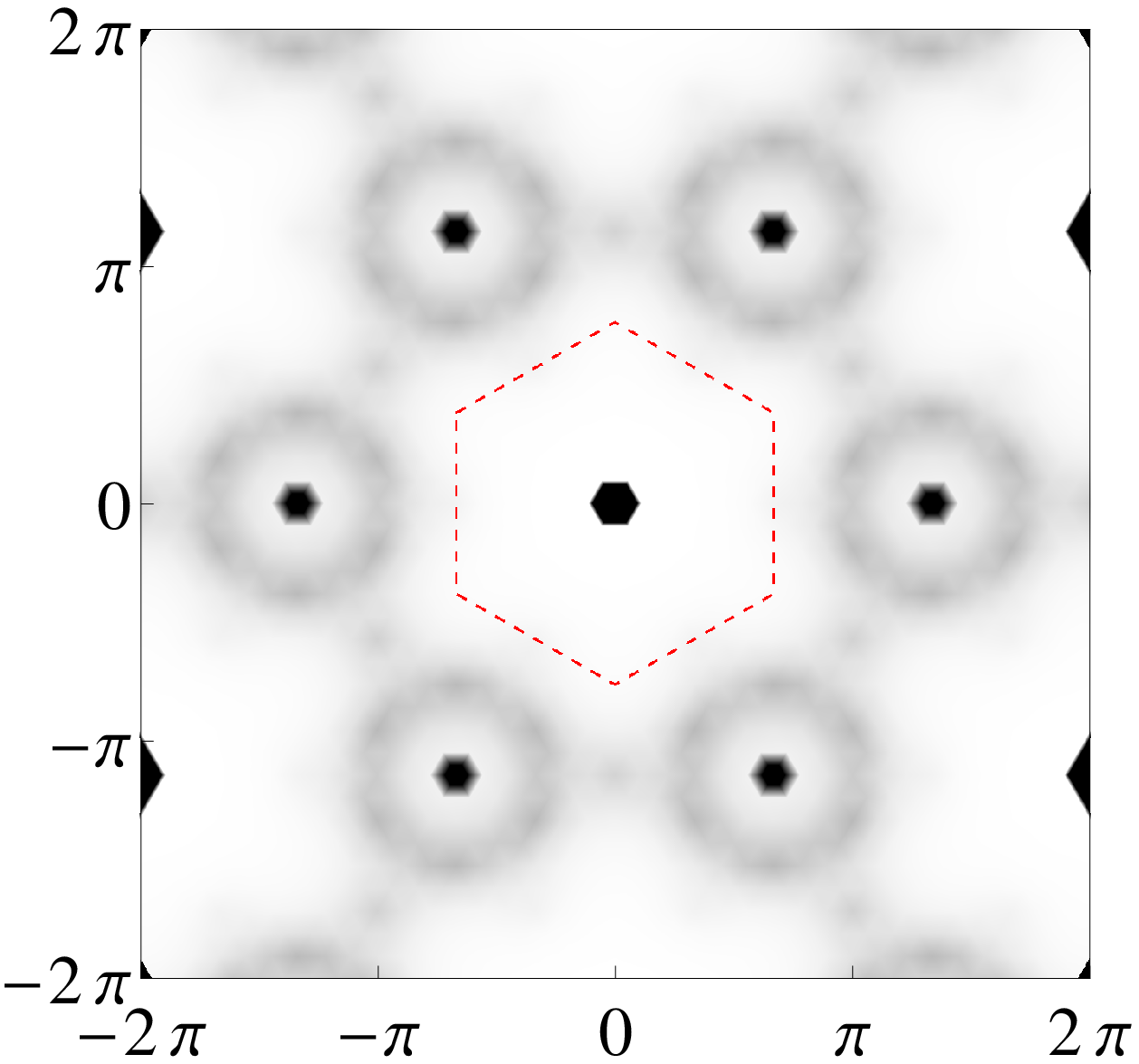}
\caption{(Color online) Structure factor $S(\mathbf{q})$, calculated on a $L = 24$ lattice for temperatures below ($T = 0.2W$, left) and above the phase transition ($T = 0.35W$, right) to the $n=5/8$ phase at $\mu=3.4W$. The dashed red hexagons indicate the first Brillouin zone.}
\label{fig:mcsf58}
\end{figure}

\begin{figure}[t]
\centering
\includegraphics[width=\columnwidth]{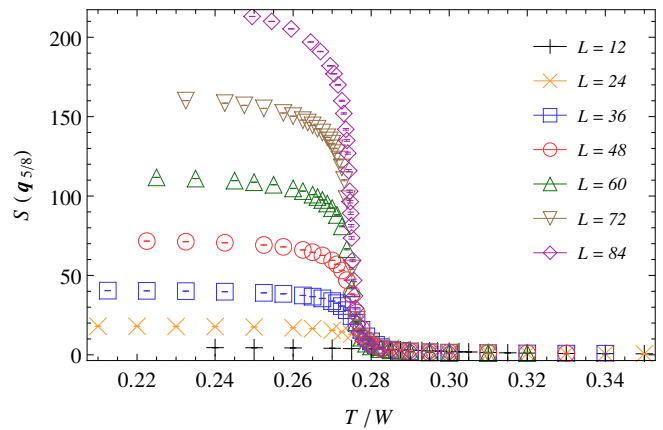}
\caption{(Color online) Structure factor $S(\mathbf{q}_{5/8})$ as a function of temperature $T$ at $\mu=3.4W$ in the  transition region to the $n=5/8$ phase for various lattice sizes.}
\label{fig:sf58}
\end{figure}

\begin{figure}[t]
\centering
\includegraphics[width=\columnwidth]{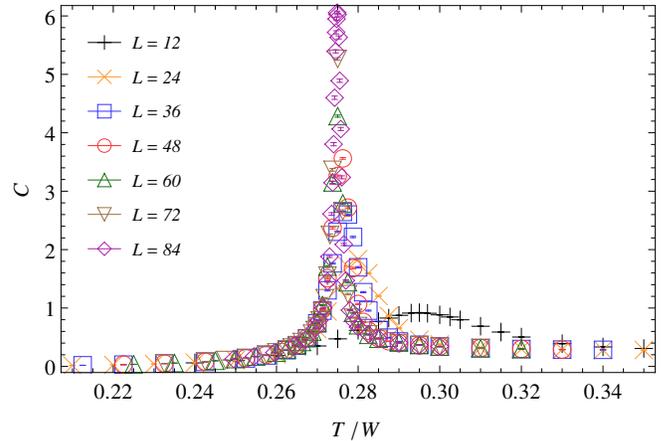}
\caption{(Color online) Specific heat $C$ as a function of temperature $T$ at $\mu=3.4W$ in the  transition region to the $n=5/8$ phase for various lattice sizes.}
\label{fig:C58}
\end{figure}

\begin{figure}[t]
\centering
\includegraphics[width=\columnwidth]{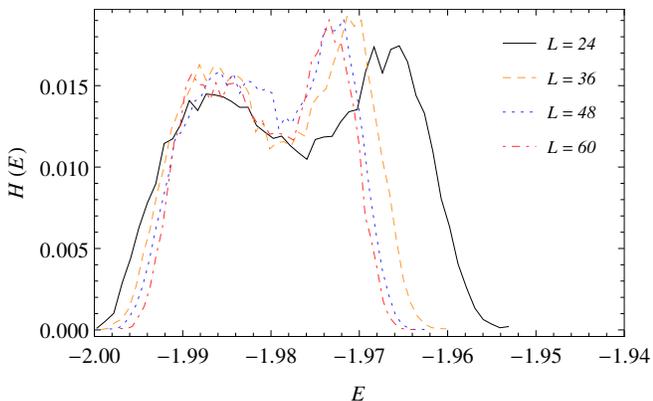}
\caption{(Color online) Histogram of the internal energy per site $H(E)$  near the transition point to the $n=5/8$ phase for various lattice sizes at $\mu=3.4W$. For each system size, $H(E)$ is shown at the temperature for which the high- and the low-energy peak exhibit similar weights.}
\label{fig:HE58}
\end{figure}

We next turn to the $n=5/8$ phase.  Here, we considered finite systems with periodic boundary conditions up to $L=84$. In the following, we present in detail our data for a chemical potential of $\mu=3.4W$.
In Fig.~\ref{fig:mcsf58}, the structure factor $S(\mathbf{q})$ is shown for different values of the temperature $T$, exhibiting at low $T$ the formation of the characteristic structure that is also seen in Fig.~\ref{fig:psf58} for the ground state. In more detail, the temperature dependence of the structure factor at the ordering wave vector, $S(\mathbf{q}_{5/8})$, is shown in Fig.~\ref{fig:sf58}.
From this data, the onset of the low-$T$ ordered phase near $T=0.275W$ can be extracted. The transition is again also monitored by the specific heat $C$, shown in Fig.~\ref{fig:C58}, which exhibits a pronounced peak near $T= 0.275W$. 
\begin{figure}[t]
\centering 
\includegraphics[width=\columnwidth]{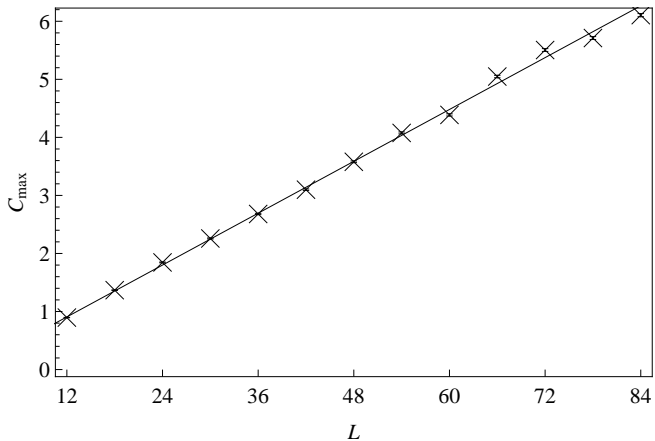}
\caption{Specific heat maximum $C_{max}$ as a function of $L$ at $\mu=3.4W$ in the transition region to the $n=5/8$ phase. The line denotes a linear increase with $L$.}
\label{fig:Cmax58}
\end{figure}
The finite-size scaling  of the specific heat maximum $C_{max}$ with the linear system size $L$ is shown in Fig.~\ref{fig:Cmax58}. 
Instead of a quadratic increase with $L$, indicative of a first-order transition, as observed for $n=9/16$, we here observe a linear increase with $L$. Further support against a first-order melting of the $n=5/8$ phase comes from analyzing  the energy histogram $H(E)$. The histograms shown in Fig.~\ref{fig:HE58} have again been  obtained by a histogram reweighing procedure, adapting for each system size the temperature such that the relative weight of the low- and the high-energy peak are of similar magnitude. 
In the present case, the peaks in the histograms tend closer towards each other upon increasing the system size. Even more prominent is the fact, that the histogram weight between the peak positions does not show a growing suppression upon increasing $L$, as would be expected for a first-order transition. 

\begin{figure}[t]
\centering
\includegraphics[width=\columnwidth]{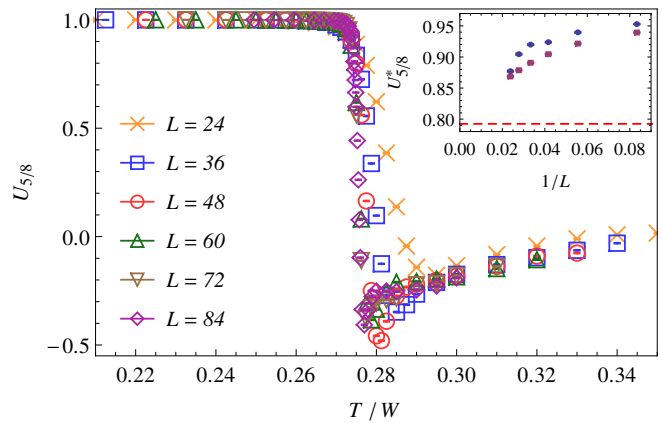}
\caption{(Color online) Binder ratio $U_{5/8}$ as a function of temperature $T$ at $\mu=3.5W$ in the  transition region to the $n=5/8$ phase for various lattice sizes. The inset shows the finite size scaling of the Binder ratio $U^*_{5/8}$ at the crossings points of $U_{5/8}$ for system sizes $L$ and $2L$ as a function of $L$ (circles), as well as the finite size values of $U_{5/8}$ at $T=0.2743$, the critical temperature obtained from a data collapse (see text) of the structure factor data (squares). The dashed line indicates the extrapolated value (in the thermodynamic limit) of the Binder ratio at the critical temperature for the four-states Potts model given in Ref.~\onlinecite{jin12}.}
\label{fig:U58}
\end{figure}

We also calculated an appropriate Binder cumulant ratio $U_{5/8}$, defined corresponding to Eq.~(\ref{eq:binder}) in terms of the structure factor $S(\mathbf{q}_{5/8})$.
The finite-size data for $U_{5/8}$ in the transition region is shown in Fig.~\ref{fig:U58}.
We again observe a dip in the Binder ratio; however in the present case, its magnitude does not increase with the system size. 
This behavior of $U_{5/8}$ is  similar to the behavior of the Binder ratio at the two-dimensional ferromagnetic four-states Potts-model transition, cf. the data in Ref.~\onlinecite{jin12}, which exhibits a continuous phase transition. 

\begin{figure}[t]
\centering
\includegraphics[width=\columnwidth]{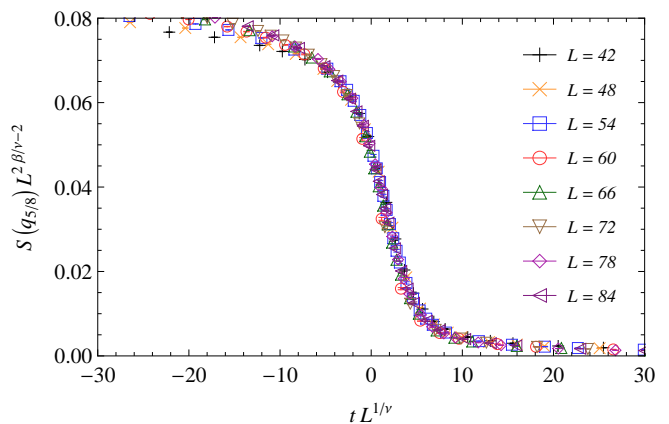}
\caption{(Color online) Data collapse plot of the structure factor data for various system sizes and temperatures at $\mu=3.5W$ in the  transition region to the $n=5/8$ phase based on two-dimensional ferromagnetic four-states Potts-model criticality.}
\label{fig:collapse58}
\end{figure}

Excluding a first-order transition, we next attempt to analyze the nature of the phase transition in terms of critical exponents of a continuous transition. The linear scaling of $C_{max}$ with the linear system size $L$ relates, via the general finite-size scaling behavior $C_{max}\propto L^{\alpha/\nu}$, to an exponent ratio of $\alpha/\nu=1$. By means of the hyper-scaling relation $d\nu=2-\alpha$, with $d=2$ the spatial dimension, this implies that 
\begin{equation}
 \alpha=\nu=\frac{2}{3}.
\end{equation}
These critical exponents relate in particular to the continuous four-states Potts-model~\cite{wu82} transition in $d=2$, which furthermore would imply a value of $\beta=1/12$. In order to assess, if these numbers are in accord also with the behavior of the spatial correlations at the phase transition, we next attempt to perform a data collapse of the finite-size structure factor data shown in Fig.~\ref{fig:sf58} to the standard finite-size scaling form
\begin{equation}
S(\mathbf{q}_{5/8})\propto L^{d-2\beta/\nu} g(tL^{1/\nu}), \quad t=\frac{T-T_c}{T_c} 
\end{equation}
with a scaling function $g$ and $T_c$ the transition temperature. A plot of the resulting data collapse with the best fit value for $T_c=0.2743$ is shown in Fig.~\ref{fig:collapse58}, displaying a rather good fit of the numerical data to the Potts-model scenario. Within the given range of system sizes, we do not obtain evidence for the presence of logarithmic factors to the dominant power-law scaling behavior, which are generically present at a four-states Potts-model-type transition in two dimensions~\cite{nauenberg79,wu82}. It thus appears possible, that similar to the Baxter-Wu model, no such logarithmic factors are present at the transition considered here.  
We finally re-examine the Binder cumulant ratio $U_{5/8}$ and consider its value at the critical point. 
From the inset of Fig.~\ref{fig:U58}, we find that the crossing point values of $U_{5/8}$ for system sizes $L$ and $2L$ exhibit a finite-size dependence that is consistent with a (slow) convergence towards the extrapolated value (in the thermodynamic limit) of the critical Binder cumulant ratio of  the four-states Potts model, given in Ref.~\onlinecite{jin12}. Also shown in the inset of Fig.~\ref{fig:U58} are the finite-size values of $U_{5/8}$ evaluated for $T=T_c$, which exhibit a similar system-size dependence.

The emergence of four-states Potts-model criticality upon entering the filling $5/8$ phase can be directly linked to the four possible ways of embedding the honeycomb super-lattice onto the underlying honeycomb lattice when constructing the ground state configuration, as discussed in Sec.~II.B. Indeed, upon cooling the system from the high-temperature phase, the spontaneous symmetry breaking exhibited by the structure factor relates to the selection of one of the four possible embeddings of the honeycomb super-lattice. Nevertheless, after selecting the embedding, the system  still explores a macroscopic number of configurations even down to the ground state at zero temperature.  
In this respect, the four-fold symmetry breaking in the present model is different from the symmetry breaking in the Potts-model or the Baxter-Wu model, where a given low-energy sector relates  to a unique ground state. 

\subsection{Filling $3/4$ Phase}
\begin{figure}[t]
\centering
\includegraphics[width=\columnwidth]{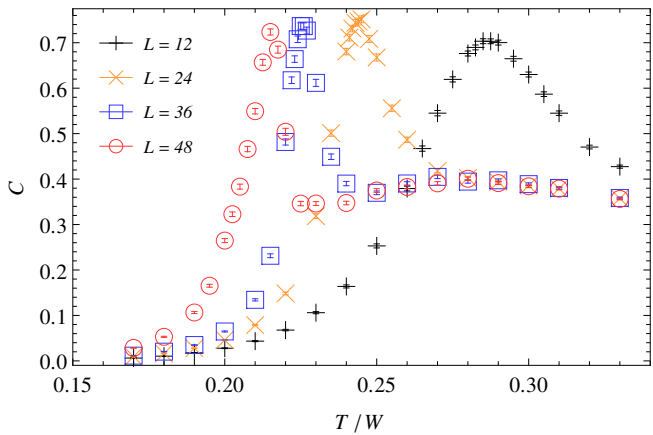}
\caption{(Color online) Specific heat $C$ as a function of temperature $T$ at $\mu=7.5W$, within the regime of the $n=3/4$ phase, for various lattice sizes.}
\label{fig:C34}
\end{figure}

As already mentioned above, we were not able to systematically study the thermal phase transition to the filling $3/4$ phase, because of  pronounced finite-size effects. These are seen e.g. in the specific heat data taken at $\mu=7.5W$, shown in Fig.~\ref{fig:C34}. 
\begin{figure}[t!]
\centering
\includegraphics[width=0.45\columnwidth]{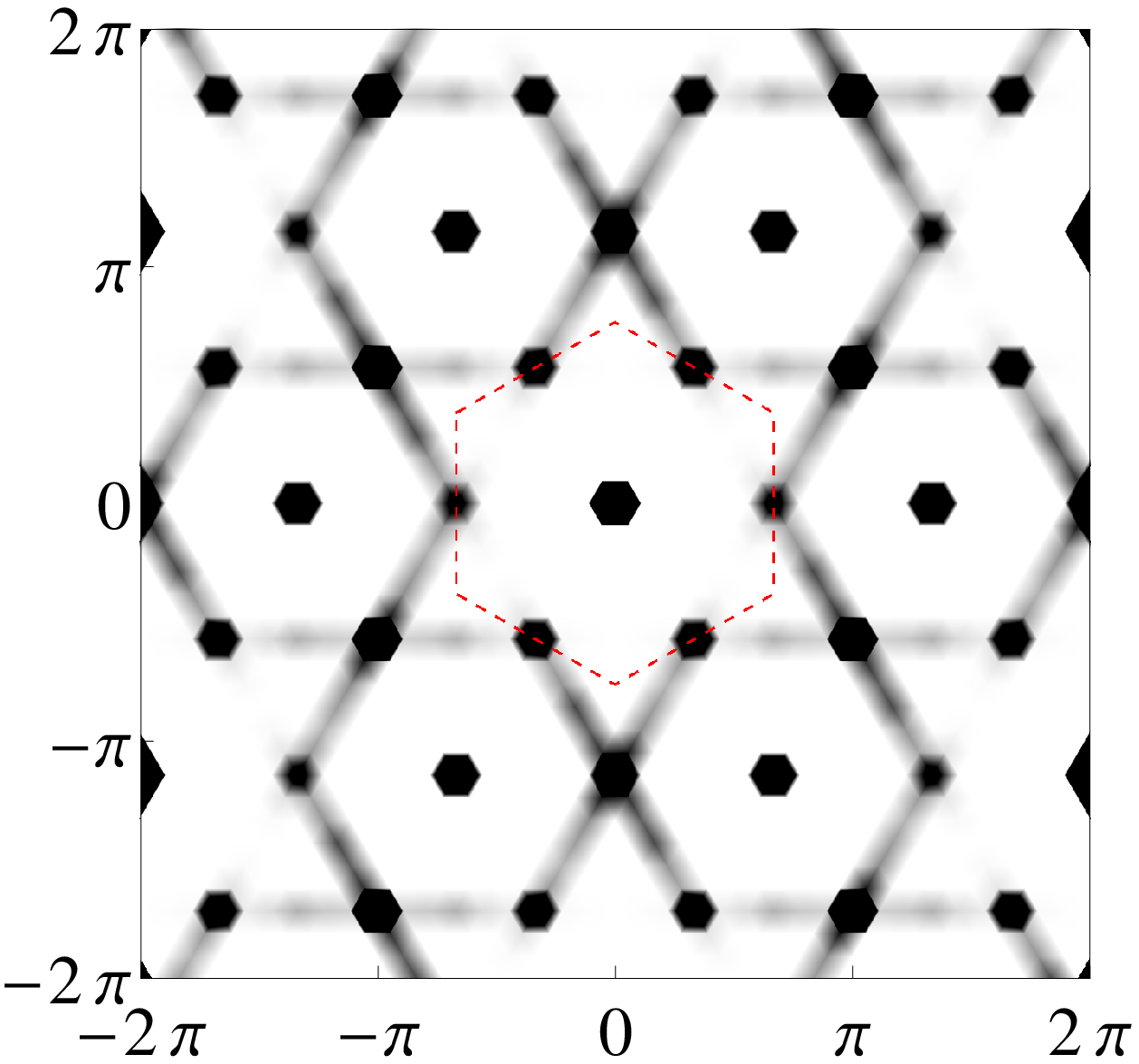}
\includegraphics[width=0.45\columnwidth]{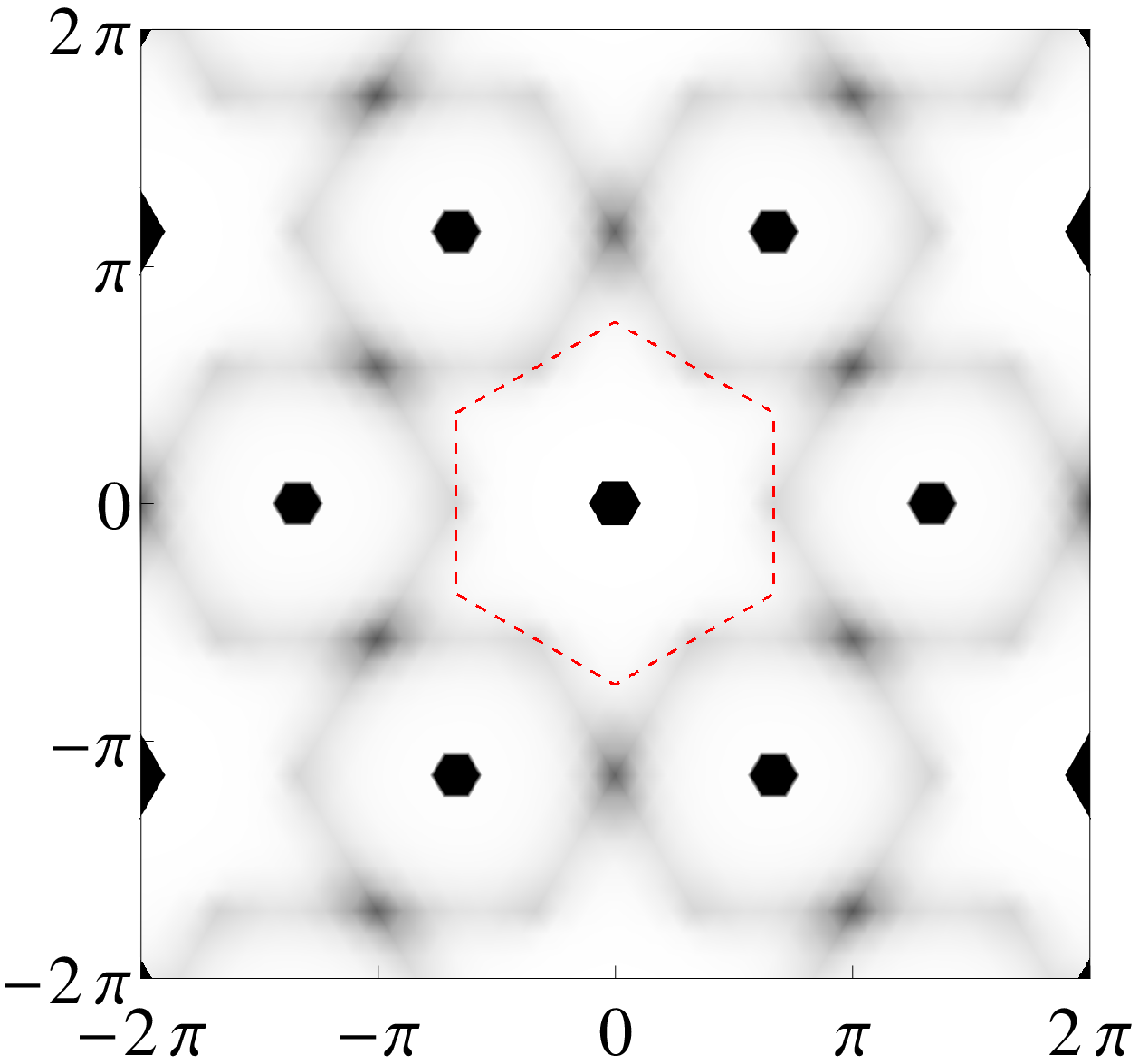}
\caption{(Color online) Structure factor $S(\mathbf{q})$, calculated on a $L = 24$ lattice for temperatures $T = 0.17W$ (left) and $T = W$ (right) for the $n=3/4$ phase at $\mu=7.5W$. The dashed red hexagons indicate the first Brillouin zone.}
\label{fig:mcsf34}
\end{figure}
We observe a strong shift of the peak position with system size, while the maximum value does not exhibit a strong finite size dependence. For $L>48$, we  were not able to obtain Monte Carlo data that did not suffer from a strong dependence on the temperature mesh employed in the parallel tempering simulations. 
However, an analysis on larger system sizes seems to be necessary in order to clarify the nature of this transition.  
Hence, we restrict here to display in Fig.~\ref{fig:mcsf34} the evolution of the  structure factor upon cooling the system at $\mu=7.5W$ on a $L=24$ lattice. At low temperatures, the characteristic structure of the $n=3/4$ phase emerges that was obtained also for the $n=3/4$ ground state (cf. Fig.~\ref{fig:psf34}).
We implemented also specific non-local updated for this regime, which attempt to shift whole lines of particles, in accord with the nature of the low-temperature phase, but they do not significantly enhance the update dynamics in the transition region.

\section{Conclusions}
We employed parallel tempering Monte Carlo simulations to access the finite temperature properties of a classical hard-core lattice gas model with three-body interactions on the honeycomb lattice that was found previously to exhibit several complex ground state configurations. We were able to identify for all three phases the onset of the corresponding structural  order at finite temperatures. For the filling $9/16$ phase, the thermal phase transition was shown to be of first-order. On the other hand, for the filling $5/8$ phase, the emergence of four-states Potts-model criticality, suggested by the numerical data, can be understood as a four-fold symmetry breaking transition related to the selection of one out of four possible embeddings of a honeycomb super-lattice. Below the transition temperature, the system still exhibits an extensive entropy all the way down to  zero temperature. It would be interesting to rationalize the apparent absence of logarithmic corrections to the dominant power-law scaling in the current model, and to assess, if e.g. a direct mapping can be established to the Baxter-Wu model, which, apart from the extensive ground state entropy, exhibits a similar four-states Potts-model criticality. For the future, it would be also interesting to explore the thermal phase transition out of the filling $3/4$ phase.

\begin{acknowledgments}
Financial support by the DFG under Grant WE 3649/2-1 is gratefully acknowledged, as well as
the allocation of CPU time within JARA-HPC and from JSC J\"ulich.
\end{acknowledgments}


%
%
\end{document}